# SUPERCONDUCTIVITY OF VARIOUS BORIDES: THE ROLE OF STRETCHED *c*- PARAMETER

Monika Mudgel, V. P. S. Awana[$] and H. Kishan
National Physical Laboratory, Dr. K. S. Krishnan Marg, New Delhi-110012, India

I. Felner
Racah Institute of Physics, Hebrew University of Jerusalem, Jeursalem-91904, Israel

Dr. G. A. Alvarez[*]
Institute for Superconducting and Electronic Materials, University of Wollongong, Australia

G. L. Bhalla
Department of Physics and Astrophysics, University of Delhi, Delhi-110007, India

[*]Presenting Author:

Dr. G. A. Alvarez
Institute for Superconducting and Electronic Materials
University of Wollongong, Australia

[$]Corresponding Author:

Dr. V.P.S. Awana

Room 109,

National Physical Laboratory, Dr. K.S. Krishnan Marg, New Delhi-110012, India

Fax No. 0091-11-45609310: Phone no. 0091-11-45609210

e-mail-awana@mail.nplindia.ernet.in: www.freewebs.com/vpsawana/

**Abstract**

The superconductivity of $MgB_2$, $AlB_2$, $NbB_{2+x}$ and $TaB_{2+x}$ is inter-compared. The stretched *c*-lattice parameter ($c$ = 3.52 Å) of $MgB_2$ in comparison to $NbB_{2.4}$ ($c$ = 3.32 Å) and $AlB_2$ ($c$ = 3.25 Å) decides empirically the population of their $\pi$ and $\sigma$ bands and as a result their transition temperature, $T_c$ values respectively at 39K and 9.5K for the first two and no superconductivity for the later. Besides the electron doping from substitution of $Mg^{+2}$ by $Al^{+3}$, the stretched *c*-parameter also affects the Boron plane constructed hole type $\sigma$-band population and the contribution from Mg or Al plane electron type $\pi$ band. This turns the electron type (mainly $\pi$-band conduction) non-superconducting $AlB_2$ to hole type (mainly $\sigma$-band conduction) $MgB_2$ superconductor (39 K) as indicated by the thermoelectric power study. Keeping this strategy in mind that stretching of *c*-parameter enhances superconductivity, the $NbB_{2+x}$ and $TaB_{2+x}$ samples are studied for existence of superconductivity. The non-stoichiometry induces an increase in *c* parameter with Boron excess in both borides. Magnetization (*M-T*) and Resistivity measurements ($\rho$-*T*) in case of niobium boride samples show the absence of superconductivity in stoichiometric $NbB_2$ sample ($c$ = 3.26 Å) while a clear diamagnetic signal and a $\rho$ = 0 transition for Boron excess $NbB_{2+x}$ samples. On the other hand, superconductivity is not achieved in $TaB_{2+x}$ case. The probable reason behind is the comparatively lesser or insufficient stretching of *c*-parameter.

# 1. Introduction

The superconductivity in the family of diborides was boost up with the discovery of $MgB_2$ superconductor in 2001 by Akimitsu group [1]. The reports regarding the nature of superconductivity in this simple compound as well as on its practical application starts appearing at a very high rate since after it's discovery. Along with this, the other diborides like $TaB_2$, $NbB_2$ & $ZrB_2$ having the same $AlB_2$ type structure were also searched for superconductivity. But in the comparison of $MgB_2$, a very few reports exist on other diborides; even the existence of superconductivity is suspected in some of the diborides. For example, $ZrB_2$ is reported to have a $T_c$ of 5.5K by Gasprov et al [2], whereas Leyrovska and Leyrovski [3] report no transition. Similarly, Gasprov et al and others [2-5] have reported no observation of superconductivity in $TaB_2$ while Kackzorowski et al [6] report a transition temperature of 9.5K. The results for $NbB_2$ are even more diverse. Gasprov et al [2], Kackzorowski et al and others [6,7] report no superconductivity while many others [3, 8-12] report different values of transition temperature in the range 0.62 to 9.2K.

All above mentioned diborides have same hexagonal structures with boron honeycomb layers sand witched between the metal hexagonal layers, but the values of lattice constants differ considerably in $MgB_2$, $AlB_2$, $NbB_2$, $NbB_{2+x}$, $TaB_2$ and $TaB_{2+x}$. The superconductivity in $MgB_2$ is attributed to it's light constituents as well as to it's stretched lattice in *c* direction with *c*/*a*=1.14, the same is not true in case of $AlB_2$ where *c*/*a* = 1.08[13, 14]. Similarly the *c*/*a* values are quite less in $NbB_2$ and $TaB_2$. Band structure calculations in $MgB_2$ reveal that $T_c$ increases with increase in *c*-parameter [15]. Working on the same idea, the stoichiometric and non-stoichiometric niobium boride and tantalum boride samples are synthesized. Non-stoichiometry is created by taking boron in excess, which results in increase of *c* parameter in both $NbB_{2+x}$ and $TaB_{2+x}$. Then the samples are checked for existence of superconductivity and the systematic comparison in both $NbB_{2+x}$ and $TaB_{2+x}$ is carried out. The thermoelectric power of stoichiometric samples of $MgB_2$, $AlB_2$ and $NbB_2$ is also intercompared.

## 2. Experimental

The polycrystalline samples of $MgB_2$, $AlB_2$, $NbB_{2+x}$ (x=0.0 to0.8) and $TaB_{2+x}$ (x=0.0 to 0.8) were prepared by simple solid-state reaction route. The constituent powders of commercial Mg, Al, $NbB_2$, $TaB_2$ and boron powders were mixed homogeneously in the stoichiometric ratios by continuous grinding. The powders were then palletized and heat-treated by argon annealing method. The phase formation was checked by X-ray diffraction patterns done on Rigaku-Miniflex-II at room temperature. Rietveld analysis was done by Fullprof program-2007 so as to obtain lattice parameters. Magnetic susceptibility measurements were carried out on a SQUID magnetometer (*MPMS-XL*). Resistivity measurements were carried out by four-probe technique. Thermoelectric power measurements were taken by differential technique on a home made set up.

## 3. Results and Discussion

Fig. 1(a) shows the X-ray diffraction patterns for $MgB_2$, & $AlB_2$ while Fig. 1 (b) shows the same for $NbB_2$ $NbB_{2.4}$, $TaB_2$ and $TaB_{2.4}$ samples. In order to confirm the phase purity, Rietveld refinement is done for all the samples in the space group P6/mmm No. 191. All the characteristic peaks are obtained at their specific Bragg position. There is hardly any difference between the experimentally observed and theoretically Rietveld determined X-ray profiles except a small MgO peak in case of $MgB_2$ shown by #. The lattice parameter values are given in respective layers of Fig.1. We observe that the (002) peak shifts towards lower angle side with the boron excess in both $NbB_2$ and $TaB_2$ case, which results in increase of *c*-parameter. The lattice parameter values are determined for all synthesized samples and the systematic variation in the parameters can be seen from Table 1. There is a slight decrease in '*a*' parameter with increasing boron content in both $NbB_{2+x}$ and $TaB_{2+x}$. In case of $NbB_{2+x}$, *c*-parameter increases continuously up to x=0.4 and then saturates further with negligible up and downs but in $TaB_{2+x}$, *c*- parameter increases considerably but only up to x=0.2 sample and saturates thereafter. The structural information is in well confirmation with the literature [6,13,16-18]. Although the *a* and *c* values for $TaB_{2+x}$ samples match quantitatively with the earlier reports [5,19] but differs in respect of corresponding compositions. $MgB_2$ is found to be a superconductor with $T_c$ of about 39 K while $AlB_2$ is a non- superconductor [14,20].

Magnetization vs temperature (*M-T*) plot including both zero field cooled (ZFC) and field cooled (FC) curves is shown in the main panel of Fig. 2(a) for $NbB_{2.4}$ sample in the temperature range 5-12 K. The $NbB_{2.4}$ sample shows a clear diamagnetic signal at about 9.5 K, implying that it is a superconductor. The lower inset of Figure 2 (a) shows magnetization vs temperature curves for $NbB_2$ sample in the temperature range 5-300 K. The sample exhibits no diamagnetic signal and hence possesses no bulk superconductivity. The magnetization measurement with varying field at a fixed temperature of 5 K is also done for the superconducting $NbB_{2.4}$ sample and is shown in the upper inset. The diamagnetic signal develops with the increasing field up to 500 Oe and after that field starts penetrating in the sample thereby reducing the diamagnetic moment with further increase of field. The diamagnetic moment reduces to near about zero value at a field of about 1600Oe. Thus, $NbB_{2.4}$ is a Type-II superconductor with the $H_{c1}$ and $H_{c2}$ values as 500 & 1600 Oe respectively. In this way Boron excess increases the *c* parameter and induces superconductivity in niobium boride sample. All boron excess samples are found to possess superconductivity with different $T_c$ values [21].

The Resistivity vs temperature measurement ($\rho$-$T$) is also carried out for both the NbB$_2$ and NbB$_{2.4}$ sample in order to confirm the existence of superconductivity. The main panel of Fig. 2(b) shows the $\rho$-$T$ measurement for NbB$_{2.4}$ sample while the inset shows the same for NbB$_2$ sample. The NbB$_{2.4}$ sample shows a sharp transition with a $T_c$ onset of 7.5 K. On the other hand the stoichiometric NbB$_2$ sample just shows metallic behavior from 300 K to about $T$=80 K. After that resistivity becomes almost constant and shows no superconducting transition down to 5 K. Thus $\rho$-$T$ measurement is in confirmation with the $M$-$T$ measurement showing that only Boron excess sample is superconducting while the stochiometric NbB$_2$ is a non-superconductor although $T_c$ onset obtained from magnetization measurement for NbB$_{2.4}$ is comparatively higher.

After inducing superconductivity in NbB$_{2+x}$, the same is tried for TaB$_{2+x}$ sample. We have seen through X-ray diffraction patterns in Fig. 1(b) that the Boron excess in TaB$_2$ also results in increase of $c$-parameter. The Magnetization vs temperature measurements ($M$-$T$) are shown in Fig. 3 for TaB$_{2+x}$ samples in the temperature range of 5 to 20 K. The samples do not exhibit any diamagnetic signal confirming that there is no superconductivity below to 5 K. The magnetic moment increases with the decrease in temperature for all the samples. The inset shows the magnetic behavior of TaB$_{2.4}$ and TaB$_{2.6}$ samples with varying field at a fixed temperature of 5 K. The magnetic moment increases with the applied field and then saturates at a field of about 4 kOe and a hysteresis is obtained in decreasing direction of field. In this way, a paramagnetic type behavior is shown by both the samples. The magnetic moment of TaB$_{2.6}$ sample is more than the TaB$_{2.4}$ sample at a particular field value, which might be due to some magnetic impurity in the boron powder.

Now the point to be discussed is that if increase in $c$-parameter induces superconductivity in NbB$_{2+x}$, why it does not happen in TaB$_{2+x}$ case. Actually, if we see the values of $c$-parameter in $NbB_{2+x}$ case, it has increased from 3.264 Å for pure $NbB_2$ to 3.320 Å for $NbB_{2.4}$ and saturates thereafter. For $TaB_2$, *c*-parameter is 3.238 Å, which is less than that of pure $NbB_2$. With boron excess, *c* parameter increases in $TaB_{2+x}$ case also but slightly i.e. only up to 3.278 Å for $TaB_{2.2}$. After that, no increase in *c*-parameter is noticed, which implies excess boron cannot be accommodated in the $TaB_2$ lattice after this limit. Excess boron actually creates metal vacancies in the system as discussed in many theoretical studies [22, 23]. So, we come to the conclusion that although *c*-parameter increases in $TaB_{2+x}$ case but it is not sufficient to create enough metal vacancies to introduce superconductivity in this system.

Fig. 4 shows the variation in thermoelectric power (TEP) of $MgB_2$, $AlB_2$ & $NbB_2$ samples with temperature. As indicated by the sign of TEP, $MgB_2$ is a hole type conductor and exhibits superconductivity at 39 K with it's Seeback co-efficient, S = 0 below this temperature, while $AlB_2$ and $NbB_2$ have electron type conductivity and are non-superconductors. The $Al^{+3}/Nb^{+5}$ substitution at $Mg^{2+}$ provides extra electrons and hence filling of the hole type sigma band and resulting electron type conductivity. As mentioned before the non-superconductiing behavior of $NbB_2$ and $AlB_2$ is seemingly due to two facts i.e., changed carrier density and the *c*-parameters. The detailed analysis of TEP data on the basis of two-band model is done earlier for $MgB_2$ and $AlB_2$ [14]. It is discussed theoretically that the presence of vacancies in the Niobium sub-lattice of $NbB_2$ brings about considerable changes in the density of states in the near Fermi region and hence affects the superconductivity [24].

## 4. Conclusion

The impact of *c* parameter is seen on the superconductivity of different diborides. The *c*-parameter is stretched for non-stoichiometric $NbB_{2+x}$ and $TaB_{2+x}$ samples. Excess boron creates metal vacancy in the lattice and induces superconductivity in niobium boride case but the increase in *c*-parameter is not sufficient in TaB$_2$ case and hence the superconductivity is not achieved. The thermoelectric power measurement shows the different type of carriers in different borides. It is concluded that the non-superconductiing behavior of NbB$_2$ and AlB$_2$ is seemingly due to two facts i.e., changed carrier density and *c*-parameters.

## 5. Acknowledgement

The authors from *NPL* would like to thank Dr. Vikram Kumar (Director, *NPL*) for showing his keen interest in the present work. One of us Monika Mudgel would also thank *CSIR* for financial support by providing *JRF* fellowship. Dr. V. Ganesan is acknowledged for thermoelectric power measurements.

## References

1. J. Nagamatsu, N. Nakagawa, T. Muranaka, Y. Zenitani and J. Akimitsu, Nature **410**, 63 (2001).

2. V. A. Gasprov, N. S. Sidorov, I. I. Zever'kova, M. P. Kulakov, JETP Lett. **73**, 601 (2001).

3. L. leyarovska, E. Leyarovski, J. Less-Common Met. **67**, 249 (1979).

4. H. Rosner, W. E. Pickett, S. L. Drechsler, A. Handstein, G. Behr, G. Fuchs, K. Nankov, K. H. Muller, H. Eshrig, Phys. Rev. B **64**, 144516 (2001)

5. A. Yamamoto, C. Takao, T. Matsui, M. Izumi, S. Tajima, Physica C **383**,197 (2002).

6. D. Kaczorowski, A. J. Zaleski, O. J. Zogal, J. Klamut, cond-mat **0103571** (2001)

7. Cooper et al, Proc. Nat. Acad. Sci. **67**, 313 (1970).

8. H. Kotegawa, K. Ishida, Y. Kitaoka, T. Muranaka, H. Takagiwa, J. Akimitsu, Physica C **378**, 25 (2002).

9. J. E. Schriber, D. L. Overmeyer, B. Morosin, E. L. Venturini, R. Baughman, D. Emin, H. Klesnar, T. Aselage, Phys. Rev. B **45**, 10787 (1992).

10. J. K. Hulm, B. T. Mathias, Phys. Rev. B **82**, 273 (1951).

11. W. A. Zeigler, R. Young, Phys. Rev. B **94**, 115 (1953).

12. H. Takeya, K. Togano, Y. S. Sung, T. Mochiku, K. Hirata, Physica C **408**, 144 (2004).

13. I. Loa, K. Kunc, K. Syaseen & P. Bouvier, Phys. Rev. B **66**, 134101 (2002).

14. Monika Mudgel, V.P.S. Awana, G. L. Bhalla, H. Kishan, L.S. Sharath Chandra, V. Ganesan, and A.V. Narlikar, J. Phys. Cond. Matt. **20**, 095205 (2008).

15. Xiangang Wan, Jinming Dong, Hongming Weng, and D. Y. Xing, Phys. Rev. B **65**, 012502 (2002).

16. A. Yamamoto, C. Takao, T. Matsui, M. Izumi, S. Tajima, Physica C **383**, 197 (2002).

17. Xiangang Wan, Jinming Dong, Hongming Weng, and D. Y. Xing, Phys. Rev. B **65**, 012502 (2002).

18. R. Escamilla, O. Lovera, T. Akachi, A. Duran, R. Falconi, F. Morales and R. Escudero, J. Phys.: Condens. Matter **16**, 5979 (2004).

19. H. Itoh, Y. Satoh, S. Kodama, S. Naka, J. Jpn. Ceram. Soc. **98**, 264 (1990).

20. V. P. S. Awana, Arpita Vajpayee, Monika Mudgel, V. Ganesan, A.M. Awasthi, G.L. Bhalla, and H. Kishan, Eur. Phys. J. B **62**, 281 (2008).

21. Monika Mudgel, V.P.S. Awana, G. L. Bhalla and H. Kishan, Solid State Communications **147**, 439 (2008).

22. L. E. Muzzy, M. Avdeev, G. Lawes, M. K. Hass, H. W. Zandbergan, A. P. Ramirez, J. D. Jorgensen, R. J. Cava, Physica C **382**, 153 (2002).

23. H. Klesner, T. L. Aselage, B. Morosin, and G. H. Kwei, J. Alloys and compounds **241**, 180 (1996).

24. I. R. Shein, N. I. Medvedeva and A. L. Ivanovskii, Physics of the solid state, **45,** 1617 (2003)

**Table 1**: Lattice parameters and *c/a* values for $NbB_{2+x}$ & $TaB_{2+x}$ samples with x = 0.0, 0.2, 0.4, 0.6 & 0.8.

| | $TaB_{2+x}$ | | | $NbB_{2+x}$ | | |
|---|---|---|---|---|---|---|
| x | *a* (Å) | *c* (Å) | *c/a* | *a* (Å) | *c* (Å) | *c/a* |
| 0.0 | 3.0899(1) | 3.2378(2) | 1.048 | 3.1103 (1) | 3.2640(2) | 1.049 |
| 0.2 | 3.0739 (1) | 3.2776(2) | 1.066 | 3.1013 (1) | 3.3051(2) | 1.066 |
| 0.4 | 3.0732 (1) | 3.2762 (2) | 1.066 | 3.1041 (2) | 3.3202(1) | 1.069 |
| 0.6 | 3.0741(1) | 3.2775(1) | 1.066 | 3.1018 (1) | 3.3195 (1) | 1.070 |
| 0.8 | 3.0746(1) | 3.2771(2) | 1.066 | 3.1040 (2) | 3.3172(2) | 1.069 |

**Figure Captions**

Fig. 1 Rietveld refined plots for (a) $MgB_2$ & $AlB_2$ samples and (b) $NbB_2$, $NbB_{2.4}$, $TaB_2$ & $TaB_{2.4}$ samples. X-ray experimental diagram (dots), calculated pattern (continuous line), difference (lower continuous line) and calculated Bragg position (vertical lines in middle).

Fig. 2(a) The magnetization vs temperature (*M-T*) plot for super conducting $NbB_{2.4}$ The lower inset shows the same for $NbB_2$ while upper inset shows the *M-H* plot for $NbB_{2.4}$ sample.

Fig. 2(b) Variation of Resistivity with temperature for $NbB_{2.4}$ sample. The inset shows the same for $NbB_2$.

Fig. 3 The *M-T* plot for $TaB_{2+x}$ sample with x = 0.0, 0.2, 0.4 & 0.6. The inset shows the *M-H* plot for $TaB_{2.4}$ & $TaB_{2.6}$ samples.

Fig. 4 Thermoelectric power vs temperature plots for $MgB_2$, $AlB_2$ & $NbB_2$ samples.

Fig. 1(a)

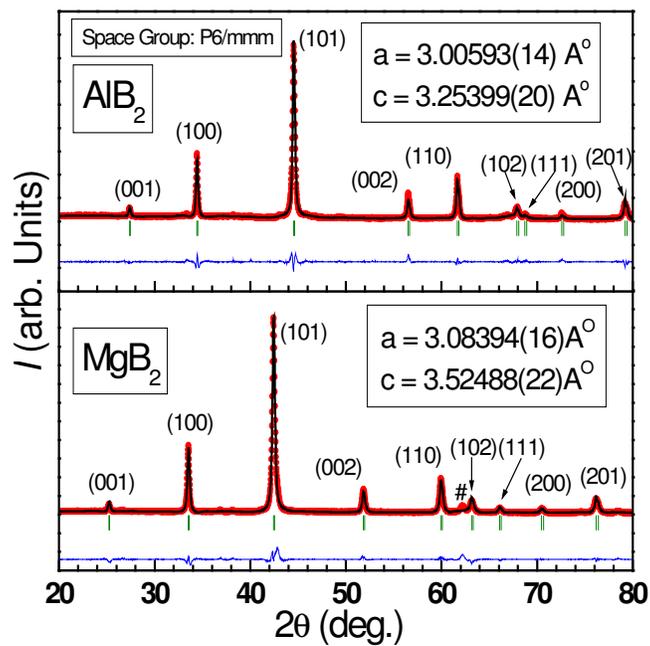

Fig. 1(b)

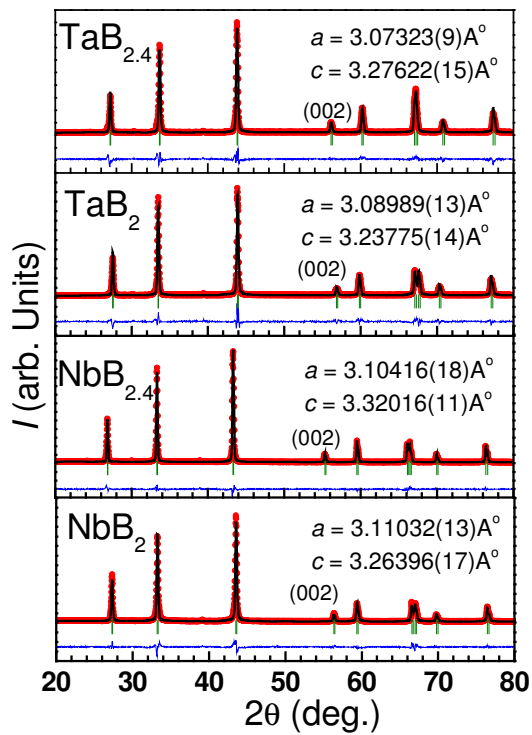

Fig. 2(a)

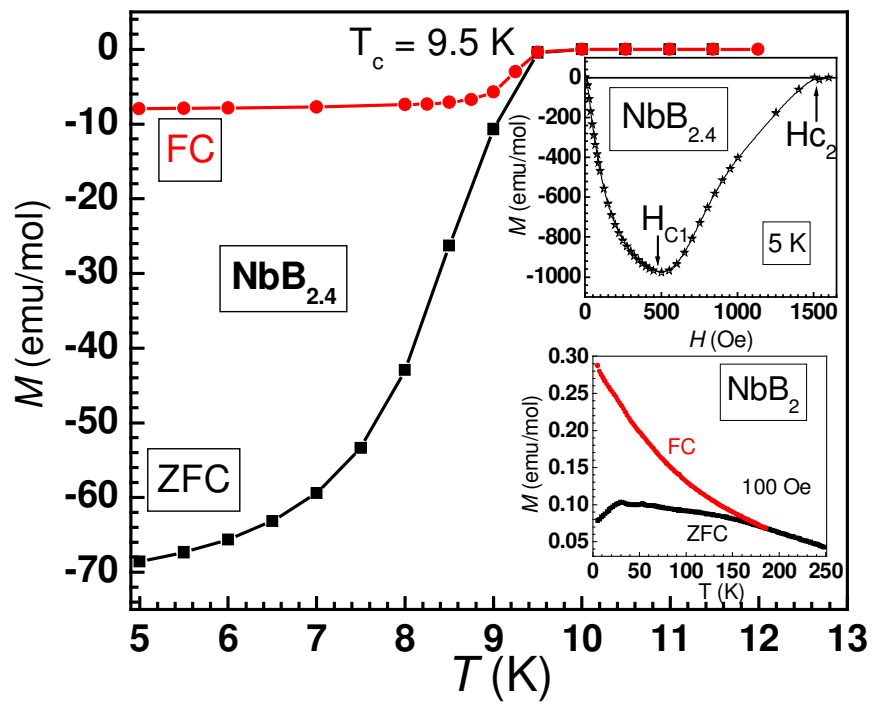

Fig. 2(b)

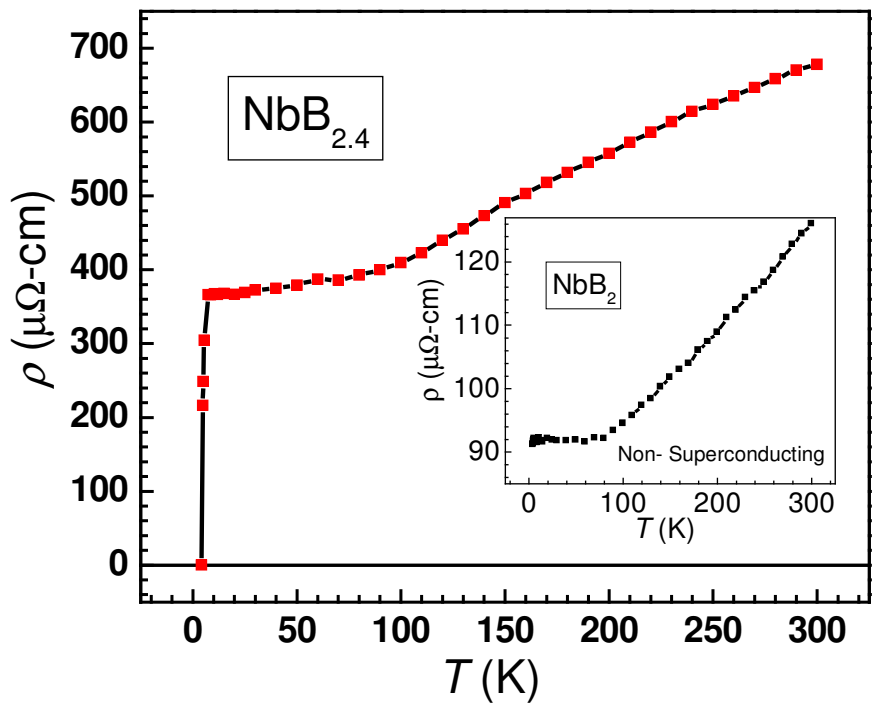

Fig.3

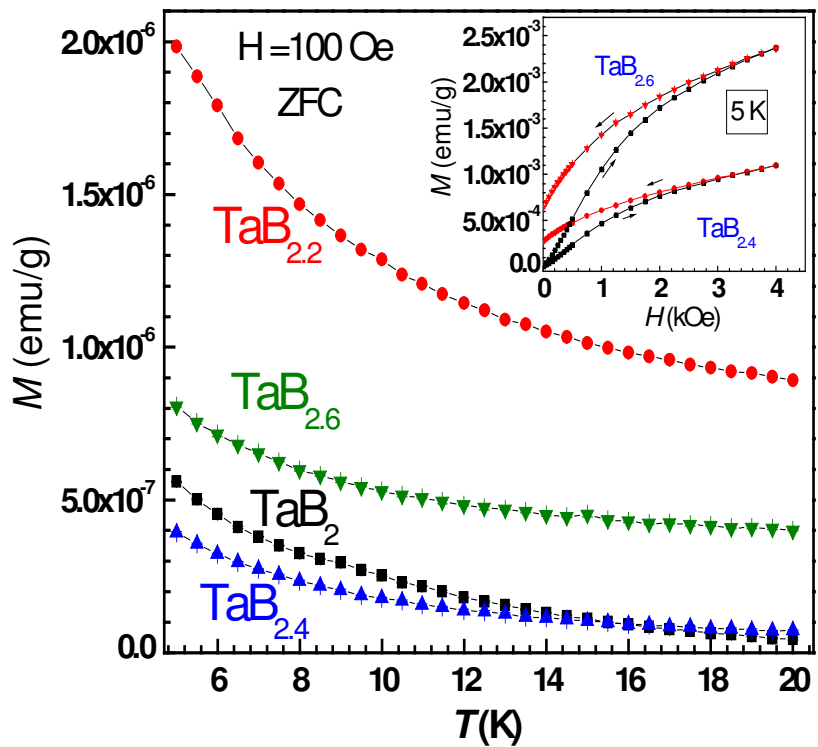

Fig. 4

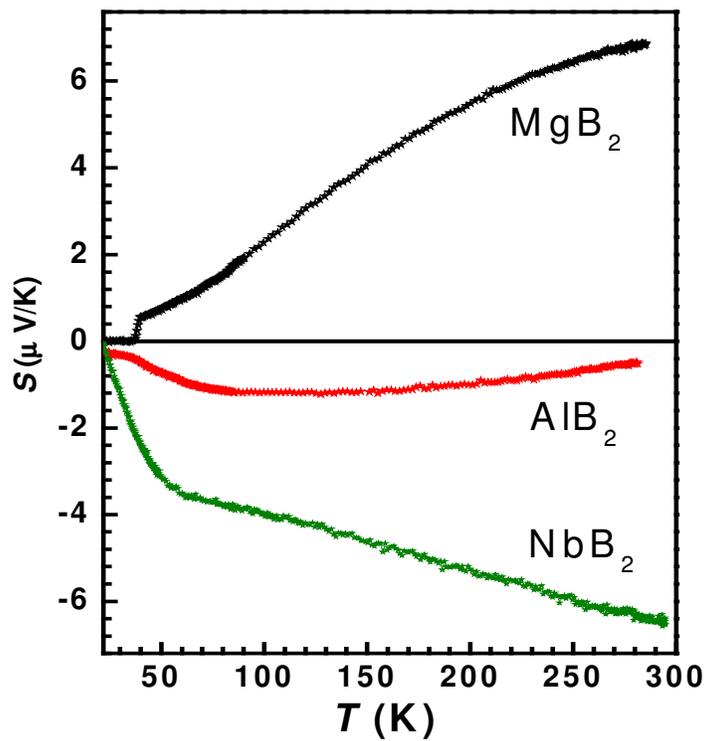